\documentstyle[12pt]{article}

\begin{document}

\begin{center}
{\Large \bf Conventional Quantum Mechanics Without Wave Function 
and Density Matrix}
\end{center}

\bigskip

\begin{center}
Vladimir I.  Man'ko 
\end{center}

\medskip

\begin{center}
{\sl
P.N. Lebedev Physical Institute, Leninskii
Pr. 53,  Moscow 117924, Russia\\
e-mail: manko@na.infn.it\\
~~~~~~~~~manko@sci.lebedev.ru}
\end{center}

\bigskip
\begin{abstract}
\noindent 
The tomographic invertable map of the Wigner function onto the positive 
probability distribution function is studied.
Alternatives to  the Schr\"odinger  evolution equation and to the energy 
level equation written for the positive probability distribution are 
discussed.  Instead of the transition probability amplitude 
(Feynman path integral) a
transition probability is introduced. 
 A new formulation of the conventional quantum mechanics (without wave
function and density matrix)  based on the
``probability representation'' of quantum states is given.
An equation for  the propagator in the new formulation of quantum mechanics 
is derived. Some paradoxes of quantum mechanics are reconsidered.  
\end{abstract}

\section*{Introduction}

\noindent

During more than 70 years  of existence of quantum mechanics, there was a 
dream to reduce misterious and intuitively very unusual notions of this 
theory to the well-known and intuitively acceptable classical notions.
There was a common prejudice that it is impossible to describe the notion of 
quantum state in the framework of the conventional quantum mechanics in terms 
of the probability density and one  is obliged to use either complex wave 
function or density matrix
 in different representations.
Fortunately, it turns out that it is possible to associate with a quantum 
state the usual probability density and  the use of the wave function or 
density matrix is not mandatory in the conventional quanrum mechanics.
This course of lectures is devoted to the new formulation of quantum mechanics.
  
Quantum mechanics is based on the concept of a complex wave 
function which satisfies the Schr\"odinger equation~\cite{Sch}. Several
attempts to give classical-like interpretations of the wave function 
were done in~\cite{Br,Mad,Bo} 
(see also~\cite{sbornik}\,). 
These attempts~\cite{Br,Mad,Bo} 
and related constructions of 
quasidistribution functions in the phase space of the 
system~\cite{Wig32,Hus40,Gla63,Sud63}
give the idea that for quantum mechanics it is impossible to describe
the state of the quantum system in terms of measurable positive probability
analogously to the case of classical statistical mechanics, where the state 
of the system is described by the positive probability distribution due
to the presence of classical fluctuations.

Nevertheless, it was shown 
recently
\cite{Mancini2,Mancini3,ManConf,ManLaser,Bregenz96,Bregenz97}
that in the framework of the symplectic tomography 
scheme~\cite{Mancini1,Ariano},  which generalizes the optical tomography
scheme~\cite{Ber,VogRi}, it is possible to introduce a classical-like 
description of a quantum state using the measurable positive probability
(instead of the complex probability amplitude).

This result was obtained because, in addition to considering a measured 
physical observable in a fixed reference frame in the phase space of the 
quantum system, different reference frames in the phase space were
considered. In the spirit of methodology, it is close to special relativity 
theory, where to get unusual effects due to motion with high velocities,
different reference frames connected by Lorentz transform must be used.
In the quantum case, the extra parameters distinguishing different reference
frames replace the information coded by a phase of the wave function.
This approach can be considered as introducing a new representation in 
quantum mechanics which can be called the ``probability 
representation''~\cite{Safonov1,Safonov2}.

The description of the quantum state in terms of positive probability was
obtained not only for continuous observables like the 
position~\cite{Bregenz96,Mancini1,Ariano}, 
but also for pure quantum observables like spin~\cite{Dodspin,JETF} 
(see also~\cite{Leonh}\,).

A classical formulation of quantum evolution was suggested, and for 
the marginal distribution a new quantum evolution
equation was found~\cite{Mancini2} which is an alternative to the 
time-dependent Schr\"odinger equation. 
This equation gives the classical-like description of quantum
evolution in terms of a normalized positive distribution containing
complete information on the state of the system. Examples of free
motion and some excited states of the harmonic oscillator were also
considered~\cite{Mancini2,Mancini3}. The evolution of even and odd 
coherent states~\cite{Dodonov74} of a particle in a Paul trap was 
investigated~\cite{Olga1} in the framework  of the classical-like 
description~\cite{Mancini2,Mancini3}. The even and odd coherent 
states of a trapped ion were discussed in recent 
papers~\cite{WVogel,Nieto}.
Experimentally these states were realized in~\cite{Winel,Haroche}.
A review of the metod of  integrals of motion  and its application to 
oscillator's models used in the Paul trap problem is given in 
Ref.~\cite{ELAF95}.

The aim of the course of lectures is to discuss, 
following~\cite{ManLaser,Bregenz96,Bregenz97,OLGAJRLR97},
the notion of a quantum state in the new formulation of quantum mechanics. 
We review the classical-like description 
of transition probabilities between stationary states (energy levels) 
of quantum systems and obtain analogs of the orthogonality
and the completeness relations. We show that, if the evolution equation
describing the dynamics of a quantum system is determined by the imaginary 
part of the system's potential energy considered as a function of a complex
coordinate, the energy states of the system are determined by the real
part of the potential energy. The energy levels of the harmonic oscillator
are rederived in the framework of classical-like alternatives to 
Schr\"odinger evolution and stationary equations. A new type of eigenvalue 
problem is formulated for the positive and normalized marginal distributions.

\section*{Classical Statistical Mechanics and Tomography Map}

\noindent

In quantum mechanics, the tomography methods of measuring quantum 
states~\cite{Mancini1,Ber,VogRi} gave the possibility to introduce new 
approach to the notion of quantum states~\cite{Mancini2,Mancini3,ManLaser}.
It turned out  that the tomography methods can be used in classical 
statistical mechanics~\cite{OLGAJRLR97,FEDELEPhysRevE98}.
Following ~\cite{OLGAJRLR97} we start from introducing the positive 
probability distribution function for  a state of  a classical system. 
In the course of  lectures, we will show that both classical and quantum 
states are described by the same probability distribution function (called 
the marginal distribution function).

Main expressions for marginal distributions in the optical 
tomography method~\cite{Ber,VogRi} as well as in the symplectic tomography 
method~\cite{Mancini1} are based on a theorem which connects the 
characteristic function with the probability distribution function.
This connection is valid for quantum states described by a 
density matrix~\cite{Cahill}. It is obvious that the same connection 
also exists for classical systems in the framework of classical
statistical mechanics. One can prove that the Fourier transform of the
characteristic function (calculated by means of a classical probability
distribution) is a positive distribution function. We illustrate this 
statement by an example of a one-dimensional system.

``States'' in classical statistics are described by the function
$f\left(q,\,p\right)$, which is the probability distribution function in 
the phase
space, i.e., 
$$f\left(q,\,p\right)\geq 0\,,\qquad \int f\left(q,\,p\right)\,dp=P(q)\,,\qquad
\int f\left(q,\,p\right)\,dq=\widetilde P(p)\,,$$
with $P(q)$ and $\widetilde P(p)$ probability distributions for position 
and momentum, respectively, ({\it marginals}).

Let the nonnegative function $f\left(q,\,p\right)$ be a
distribution function of the classical system in the phase space.
The coordinates $-\infty <q<\infty $ and $-\infty <p<\infty $ are
the position and momentum of the system, respectively. The function
$f\left (q,\,p\right )$ is taken to be normalized
\begin{equation}\label{csm1}
\int  f\left (q,\,p\right)\,dq\,dp=1\,.
\end{equation}
We consider an observable $X\left (q,\,p\right)$ which is a function on the 
phase space of the system under study. For the case of classical statistical
mechanics, the characteristic function for the observable 
$X\left (q,\,p\right)$ 
\begin{equation}\label{csm2}
\chi \left (k\right )=\langle e^{ikX}\rangle
\end{equation}
is given by the relation
\begin{equation}\label{csm3}
\chi \left (k\right)=\int e^{ikX(q,p)}\,f\left (q,\,p\right)
\,dq\,dp\,.
\end{equation}
The Fourier transform of the characteristic function
\begin{equation}\label{csm4}
w\left (X\right )=\frac {1}{2\,\pi}\,\int \chi \left (k\right )e^{-ikX}\,dk
\end{equation}
is a real nonnegative function which is normalized
\begin{equation}\label{csm5}
\int w\left (X\right )\,dX=1\,.
\end{equation}
In fact, due to Fourier representation of Dirac delta-function, one has
\begin{equation}\label{csm6}
w\left (X\right )=\int  f\left(q,\,p\right)\,\delta
\left (X(q,\,p)-X\right )\,dq\,dp\,.
\end{equation}
The distribution function is nonnegative and the delta-function is also
nonnegative. So we integrate the product of two nonnegative functions over
the phase space. The result of the integration $w\left (X\right )$ is
obviously a nonnegative function. 

Let us now check the normalization of the function $w\left (X\right ).$  
We have    
\begin{equation}\label{csm7}
\int w\left (X\right )\,dX=\int f\left( q,\,p\right)
\,\delta \left (X(q,\,p)-X\right )\,dq\,dp\,dX\,.
\end{equation}
In view of the definition of delta-function, one has 
\begin{equation}\label{csm8}
\int \delta \left (X(q,\,p)-X\right )\,dX=1\,.
\end{equation}
This means that 
\begin{equation}\label{csm9}
\int w\left (X\right )\,dX=\int  f\left( q,\,p\right )\,dq\,dp\,.
\end{equation}
Since the distribution function $f\left (q,\,p\right )$ 
satisfies the normalization condition~\ref{csm1}), we have shown that
the Fourier transform of the characteristic function $w\left (X\right )$
given by~\ref{csm4}) is normalized too, i.e., it satisfies the normalization
condition~\ref{csm5}). As a classical analog of the quantum 
symplectic-tomography observable introduced in Ref.~\cite{Mancini1} we consider
the classical observable which is a linear function on the phase space of the
system,
\begin{equation}\label{csm10}
X\left (q,\,p\right )=\mu q+\nu p\,,
\end{equation}
where the real parameters $\mu $ and $\nu $ are interpreted as the parameters 
of symplectic transform of the position and momentum of the system under study.
(We discuss only one variable ---the position $X\left (q,\,p\right )$---
and do not take into account the conjugate momentum.) 

The variable $X\left (q,\,p\right )$ can be considered from two equivalent 
points of view. It can be interpreted as a canonically transformed position
which is a linear combination of position and momentum in a fixed reference 
frame in the phase space of the system. Another equivalent interpretation of 
the variable $X\left (q,\,p\right )$ given by Eq.~\ref{csm10}) is that it is a 
position of the system measured in the rotated and scaled reference frame in 
the classical phase space of the system.

We use the second interpretation, according to which the real parameters
$\mu $ and $\nu $ determine the reference frame in the phase space of the
system in which the position is measured. For the position in the transformed
reference frame~\ref{csm10}), we get from Eq.~\ref{csm3}) the distribution 
function (the tomography map)
\begin{equation}\label{csm11}
w\left (X,\,\mu ,\,\nu \right )=\frac {1}{2\,\pi}\int e^{-ik(X-\mu q-\nu p)}
\,f\left( q,\,p\right)\,dq\,dp\,dk\,.
\end{equation}
Another form for the probability distribution is given by Eq.~\ref{csm6})   
\begin{equation}\label{csm12}
w\left (X,\,\mu ,\,\nu \right )=\int  f\left (q,\,p\right )\,
\delta \left (\mu q +\nu p -X\right )\,dq\,dp\,.
\end{equation}
One can see that the marginal distribution is a homogenious function, i.e., 
\begin{equation}\label{new1a}
w\left(\lambda X,\,\lambda \mu,\,\lambda \nu\right)=|\lambda|^{-1}
w\left(X,\,\mu,\,\nu\right).
\end{equation}
We introduced the notation $w\left (X,\,\mu ,\,\nu \right )$ for the probability
distribution of the position of the classical system in the transformed 
reference frame in the phase space to point out the dependence of the
distribution on the parameters $\mu $ and $\nu $ determining the reference
frame. Due to the dependence of the distribution 
$w\left (X,\,\mu ,\,\nu \right )$ on these parameters, we call the 
distribution a marginal distribution function.

The partial case of the canonical transform~\ref{csm10}) is a rotation in
the phase space
\begin{equation}\label{csm13}
X=q\,\cos \,\varphi +p\,\sin \,\varphi \,.
\end{equation}
This means that we choose the parameters of the symplectic transform
\begin{equation}\label{csm14}
\mu =\cos \,\varphi \,,\qquad \nu=\sin \,\varphi \,.
\end{equation}
By introducing the notation for the marginal distribution of the 
rotated position
\begin{equation}\label{csm15}
w\left (X,\,\varphi \right )=w\left (X,\,\mu=\cos \,\varphi ,\,
\nu =\sin \,\varphi \right ),
\end{equation}
we get, in view of \ref{csm12}),
\begin{equation}\label{csm16}
w\left (X,\,\varphi \right )=\int  f\left (q,\,p\right )\,
\delta \left (q\cos \,\varphi +p\sin \,\varphi -X\right )\,dq\,dp\,.
\end{equation}
Using Eq.~\ref{csm11}) we have another representation for the marginal
distribution of the rotated position, namely,
\begin{equation}\label{csm17}
w\left (X,\,\varphi \right )=\frac {1}{2\,\pi}\int 
e^{-ik\left(X-q\cos \,\varphi -p\sin \,\varphi \right )}\,
f\left (q,\,p\right )\,dq\,dp\,dk\,.
\end{equation}
Introducing the transformed position and momentum
\begin{equation}\label{csm18}
Q=q\cos \,\varphi +p\sin \,\varphi \,;\qquad 
P=-q\sin \,\varphi +p\cos \,\varphi \,,
\end{equation}
in view of the invariance of the volume in the phase space
\begin{equation}\label{csm19}
dq\,dp=dQ\,dP\,,
\end{equation}
we get, using the Fourier representation of delta-function,
\begin{equation}\label{csm20}
w\left (X,\,\varphi \right )=\int \delta \left( X-Q\right )\,f
\left (Q\cos \varphi -P\sin \,\varphi ,\,Q\sin \,\varphi +P\cos \,\varphi
\right )\,dQ\,dP\,,
\end{equation}
or 
\begin{equation}\label{csm21}
w\left (X,\,\varphi \right)=\int  f\left (X\cos \,\varphi -
P\sin \,\varphi ,\,X\sin \,\varphi +P\cos \,\varphi \right )\,dP\,.
\end{equation}
Formula~\ref{csm21}) is mathematically identical to Eq.~12) 
of Ref.~\cite{VogRi} where the marginal distribution for the homodyne 
observable 
was considered. But in~\ref{csm21}), the positive classical distribution
$f\left (q,\,p\right )$ in the phase space is used instead of
the Wigner function $W\left (q,\,p\right )$ elaborated in Eq.~12) 
of Ref.~\cite{VogRi}. It is worth noting that the form of the expression
for the marginal distribution $w\left (X,\,\varphi \right)$ is invariant.
The only difference between the quantum and classical statistics in the
context of the expression for the marginal distribution 
$w\left (X,\,\varphi \right )$ is in the difference between the
classical distribution in the phase space and the Wigner function. The
Wigner function $W\left (q, \,p\right )$ can take negative values. The 
classical distribution function $f\left (q,\,p\right )$
takes only nonnegative values. Nevertheless, the result of integration in 
both cases gives the nonnegative marginal distribution 
$w\left (X,\,\varphi \right )\,.$

Formula~{\ref{csm11}) has the inverse
\begin{equation}\label{csm22}
f\left (q,\,p\right)=\frac {1}{4\,\pi^2}\int 
w\left (X,\,\mu ,\,\nu \right)\,\exp \left [-i\left (\mu q+\nu p -X\right )
\right ]\,dX\,d\mu \,d\nu \,.
\end{equation}
In classical statistical mechanics, the admissible marginal distributions 
in formula~\ref{csm22}) always satisfy the
condition that the result of  convolution 
$f\left (q,\,p\right )$ is a nonnegative function.

We have shown that instead of the distribution function 
$f\left (q,\,p\right)$ the state of the classical system in 
the framework of classical statistical mechanics can be determined by
the marginal distribution function $w\left (X,\,\mu,\,\nu \right )$, in
complete analogy with the quantum case where the symplectic tomography
procedure is used~\cite{Mancini1}. 
 Since the map
$$f\left (q,\,p\right )\Longrightarrow w\left (X,\,\mu ,\,\nu
\right )$$
is invertable, the information contained in the distribution function
$f\left (q,\,p\right )$ is equivalent to the information 
contained in the marginal distribution $w\left (X,\,\mu ,\,\nu \right ).$
For $\mu =\cos \,\varphi $ and $\nu =\sin \,\varphi ,$ we have an analog of the 
optical tomography procedure developed for the quantum case in~\cite{VogRi}. 
We have to invert formula~\ref{csm21}).
The inverse is given
by the Radon transform (see, for example, Eq.~13) in~\cite{VogRi} and
 also~\cite{wuensche}\,).
For example, if one introduces the distribution function in the form
\begin{equation}\label{1*}
f\left(q,\,p\right)=\delta \left(q-x_0\right)\delta \left(p-p_0\right),
\end{equation}
the marginal distribution takes the form 
\begin{equation}\label{2*}
w\left(X,\,\mu,\,\nu\right)=\delta \left(X-\mu x_0-\nu p_0\right)
\end{equation}
and 
\begin{equation}\label{3*}
w\left(X,\,\varphi \right)=\delta \left(X-x_0\cos \varphi -p_0\sin 
\varphi\right).
\end{equation}

For classical statistical mechanics, the tomography maps  discussed connect 
the positive distributions, and in this context our understanding of 
the notion of the classical state for systems with fluctuations is unchanged.

The evolution equation for the classical distribution function for a particle
with mass $m=1$ and potential $U(q),$
\begin{equation}\label{csm23}
\frac {\partial  f\left (q,\,p,\,t\right )}{\partial t}+
p\,\frac {\partial  f\left (q,\,p,\,t\right)}{\partial q}-
\frac {\partial U(q)}{\partial q}\,\frac {\partial  f\left (
q,\,p,\,t\right )}{\partial p}=0
\end{equation}
can be rewritten in terms of the marginal distribution $w\left (X,\,\mu ,\,
\nu ,\,t\right)$
\begin{equation}\label{csm24}
\dot w-\mu \frac {\partial}{\partial \nu }w-\frac {\partial U}{\partial q}
\left (\widetilde q\right)\left [\nu \frac {\partial }{\partial X}w\right ]
=0\,,
\end{equation}
where the argument of the function $\partial U/\partial q$ is replaced by 
the operator
\begin{equation}\label{csm25}
\widetilde q=-\left (\frac {\partial }{\partial X}\right )^{-1}
\frac {\partial }{\partial \mu }\,.
\end{equation}  
For the harmonic oscillator with frequency $\omega =1,$ the potential energy
term $U(q)=q^2/2$ gives in Eq.~\ref{csm24}) the following evolution equation
\begin{equation}\label{csm26}
\dot w-\mu \,\frac {\partial w}{\partial \nu }+\nu \,\frac {\partial w}
{\partial \mu}=0\,.
\end{equation}
We used the equality
\begin{equation}\label{csm27}
\frac {1}{2}\,\frac {\partial q^2}{\partial q}\left (\widetilde q\right )
=-\left(\frac {\partial }{\partial X}\right)^{-1}\,
\frac {\partial}{\partial \mu }
\end{equation}
and the property
\begin{equation}\label{csm28}
\left (\frac{\partial }{\partial X}\right )^{-1}\,
\frac {\partial }{\partial \mu }
\,\nu \,\frac {\partial }{\partial X}=\nu \,\frac {\partial }{\partial \mu }\,.
\end{equation}

 For the
mean value of position in classical statistics, we have
\begin{equation}\label{csm30}
\langle q\rangle =\int  f\left (q,\,p\right)q\,dq\,dp=i\,\int
w\left (X,\,\mu,\,\nu\right )e^{iX}\,\delta ^\prime \left (\mu \right)\,
\delta \left (\nu \right )\,dX\,d\mu \,d\nu \,.
\end{equation}
One can see that in classical statistical mechanics there exists a function
associated to the position, and by means of this function one can calculate
the mean value of position using the marginal distribution
$w\left (X,\,\mu ,\,\nu \right ).$

In classical statistical mechanics, one can introduce the propagator \\
$\Pi _{\mbox {cl}}
\left (X,\,\mu ,\,\nu ,\,X^\prime ,\,\mu ^\prime ,\,\nu ^\prime ,\,t_2,
\,t_1\right )$ that connects the two marginal distributions given for times
$t_1$ and $t_2\left(t_2>t_1\right )$
\begin{equation}\label{csm34}
w\left (X,\mu ,\nu ,t_2\right )=\int \Pi _{\mbox {cl}}\left (X,
\mu ,\nu ,X^\prime ,\mu ^\prime ,\nu ^\prime ,t_2,t_1\right )
w\left (X^\prime ,\mu ^\prime ,\nu ^\prime ,t_1\right )\,dX^\prime
\,d\mu ^\prime \,d\nu ^\prime .
\end{equation}
The propagator satisfies the following equation
\begin{eqnarray}\label{csm35}
&&\frac {\partial \Pi _{\mbox {cl}}}{\partial t_2}-\mu \,\frac {\partial }
{\partial \nu }\Pi_{\mbox {cl}}-\frac {\partial U(q)}{\partial q}
\left (\hat q\right )\nu \,\frac {\partial }{\partial X}
\Pi _{\mbox {cl}}\nonumber\\
&=&\delta \left (t_2-t_1\right )\delta \left (X-X^\prime \right )
\delta \left (\mu -\mu ^\prime \right )\delta \left (\nu -\nu ^\prime
\right ),
\end{eqnarray}
which follows from the evolution equation~\ref{csm24}).

Any integral of motion $I\left (q,\,p,\,t\right )$ in classical statistical
mechanics satisfies the equation
\begin{equation}\label{csm36}
\frac {dI}{dt}=\frac {\partial I}{\partial t}+\frac {\partial I}{\partial q}
\,\frac {\partial H}{\partial p}-\frac {\partial I}{\partial p}\,
\frac {\partial H}{\partial q}=0\,,
\end{equation}
where $H\left (q,\,p,\,t\right )$ is the Hamiltonian of the classical system.

Equation~\ref{csm36}) coincides for
$$H=\frac {p^2}{2}+U(q)$$
with the equation for the classical distribution function~\ref{csm23}),
\begin{equation}\label{csm37}
\frac {\partial I\left (q,\,p,\,t\right )}{\partial t}+p\,\frac {\partial
I\left (q,\,p,\,t\right )}{\partial q}-\frac {\partial U(q)}{\partial q}\,
\frac {\partial I\left (q,\,p,\,t\right )}{\partial p}=0\,.
\end{equation}
This follows from the fact that the distribution function itself is the 
integral of motion. 

If one introduces the map~\ref{csm11}) for the integrals of motion
\begin{equation}\label{csm38}
{\cal I}\left (X,\,\mu,\,\nu ,\,t\right )=\frac {1}{2\,\pi}\int
e^{-ik\left (X-\mu q-\nu p\right )}\,I\left (q,\,p,\,t\right )\,dq\,dp\,dk\,,
\end{equation}
the integral of motion ${\cal I}\left (X,\,\mu ,\,\nu ,\,t\right )$
satisfies Eq.~\ref{csm24}) in which one has to make the replacement
$w\rightarrow {\cal I}.$

In classical statistical mechanics, the distribution function 
$f\left (q,\,p,\,t\right )$ is a function of the integrals of
motion, and the propagator that determines the evolution of the distribution 
function has the form
\begin{equation}\label{csm39}
P\left (q,\,p,\,q^\prime ,\,p^\prime ,\,t\right )=
\delta \left (q^\prime -q_0\left(q,\,p,\,t\right)\right )\,
\delta \left (p^\prime -p_0\left (q,\,p,\,t\right )\right ),
\end{equation}
where $q_0\left (q,\,p,\,t\right )$ and $p_0\left (q,\,p,\,t\right )$
are integrals of motion which have the  following property:
\begin{equation}\label{csm40}
q_0\left (q,\,p,\,0\right )=q\,,\qquad p_0\left (q,\,p,\,0\right )=p\,.
\end{equation}
Using Eq.~\ref{csm39}) one can find the propagator for 
the marginal distribution function.

For example,  the initial distribution~\ref{1*}) takes the form
\begin{equation}\label{1**}
f_0\left(q,\,p,\,t\right)=\delta \left(p-p_0\right)
\delta \left(q-tp-x_0\right)
\end{equation}
and the initial marginal distribution~\ref{2*}) reads
\begin{equation}\label{2**}
w_0\left(X,\,\mu,\,\nu,\,t\right)=\delta 
\left(X-\mu tp_0-\mu x_0-\nu p_0\right).
\end{equation}
 
In quantum and classical  statistical mechanics, 
the forms of the propagators determining the evolution 
of the marginal distributions
$w\left (X,\,\mu ,\,\nu ,\,t\right )$ 
are identical for linear systems like an oscillator or free motion.

\section*{New Notion of Quantum State}

\noindent

We consider now a new  approach  to the notion  of quantum state.
It was shown~\cite{Mancini1} that for the  generic linear combination
of quadratures which is a measurable observable 
$\left (\hbar =1\right)$
\begin{equation}\label{X}
\widehat X=\mu \hat q+\nu\hat p\,,
\end{equation}
where $\hat q$ and $\hat p$ are the position and momentum, 
respectively; the marginal distribution
$w\,(X,\,\mu,\,\nu )$ (normalized with respect to the 
variable $X$), depending on the two extra real parameters
$\mu $ and $\nu ,$ is related to the state of the quantum system 
expressed in terms of its Wigner function $W(q,\,p)$ as follows:
\begin{equation}\label{w}
w\left (X,\,\mu,\,\nu \right )=\int \exp \left [-ik(X-\mu q-\nu
p)\right ]W(q,\,p)\,\frac {dk\,dq\,dp}{(2\pi)^2}\,.
\end{equation}
We use the same notation as in the classical case. If 
one has a pure state with the wave function  
$\Psi \left(y\right)$, the marginal distribution has the form 
found in Ref.~\cite{MENDES}
\begin{equation}\label{wp}
w\left(X,\,\mu,\nu\right)=\frac {1}{2\,\pi |\nu |}\left|\int\Psi\left(y\right)
\exp\left(\frac {i\mu y^2}{2\nu}-\frac {iyX}{\nu}\right)\,dy\right|^2.
\end{equation}
The physical meaning of the parameters $\mu $ and $\nu $ is that 
they describe an ensemble of rotated and scaled reference frames 
in which the position $X$ is measured. For $\mu =\cos \,\varphi $
and $\nu =\sin \,\varphi ,$ the marginal distribution~\ref{w}) is
the distribution for the homodyne-output variable used in optical
tomography~\cite{VogRi}. Formula~\ref{w}) can be inverted and 
the Wigner function of the state can be expressed in terms of the 
marginal distribution~\cite{Mancini1}\,:
\begin{equation}\label{W}
W(q,\,p)=\frac {1}{2\pi }\int w\left (X,\,\mu ,\,\nu \right )
\exp \left [-i\left (\mu q+\nu p-X\right )\right ]
\,d\mu \,d\nu \,dX\,.
\end{equation}
Since the Wigner function determines completely the quantum state of a system 
and, on the other hand, this function itself is completely determined by the 
marginal distribution, one can understand the notion of the quantum state in 
terms of the classical marginal distribution for squeezed and rotated 
quadrature. 

{\it So, we say that the quantum state is given if the position probability 
distribution $w\left(X,\,\mu,\,\nu \right)$ in an ensemble of 
rotated and squeezed reference frames in the classical phase space is 
given.}
    
It is worth noting, that the information contained in the marginal 
distribution $w\left(X,\,\mu,\,\nu \right)$ is overcomplete. 
To determine the quantum state completely, it is sufficient to give the 
function for arguments with the constraints 
$\left(\mu^2+\nu^2=1\right)$ 
which corresponds to the optical tomography scheme~\cite{VogRi,Raymer}, i.e.,
$\mu=\cos \varphi$ and the rotation angle $\varphi$ labels the 
reference frame in the classical phase space.

So, we formulate also the notion of quantum states as follows:

{\it We say that the quantum state is given if the position 
probability distribution $w\left(X,\,\varphi \right)$ in an ensemble of 
rotated reference frames in the classical phase space is given.}
    
Since the quantum state is defined by the position distribution, one could
associate an entropy with the state using the standard relation known in 
classical probability theory, i.e., the entropy 
$S\left(\mu ,\,\nu \right)$ is given by the formula
\begin{equation}\label{ent}
S\left(\mu ,\,\nu \right)=-\int \,dX\,w\left(X,\,\mu ,\,
\nu \right)\,\ln \left[w\left(X,\,\mu ,\,\nu \right)\right]\,.
\end{equation}
If we use the distribution $w\left(X,\,\varphi \right),$ the entropy 
$S\left(\varphi \right)$ depends only on the rotation angle.

The discription of quantum states by the probability function gives 
the possibility to formulate quantum mechanics without using the wave 
function or density matrix. These ingredients of the quantum theory can 
be considered as objects which are not mandatory ones since 
they are not directly measurable. The marginal probability distribution 
function 
$w\left(X,\,\mu,\,\nu\right)$,  which can be measured directly, replaces 
the wave function in the new formulation of quantum mechanics. Since 
the quantum mechanics formalism is reduced to the formalism of 
classical probability theory, well-known results of the probability 
theory can be used to get new results in quantum theory (including 
quantum computing, teleportation, and quantum cryptography).

One can also use the introduced formulation of the notion of quantum states
to describe situations in which the states are either close or essentially
different. We say that two states are close if their distributions are close,
i.e., all the highest momenta of the distributions differ very slightly. We
also say that two states are substantially different if their distributions
differ substantially, i.e., there are highest momenta for the two distributions
with large corresponding differences. 
The notion of distance in quantum mechanics using the 
tomography map was discussed in~\cite{DISTSCRIPTA}.

\section*{Quantum Evolution and Energy Levels }

\noindent

As was shown in~\cite{Mancini2}, for systems with the Hamiltonian
\begin{equation}\label{HV}
H=\frac{{\hat p}^2}{2}+V(\hat q)\,,
\end{equation}
the marginal distribution satisfies the quantum time-evolution equation, 
being the integral equation determined by the imaginary part of the
potential energy considered as a function of a complex coordinate. 
The evolution equation reads
\begin{eqnarray}\label{FPeq}
\dot w-\mu \,\frac{\partial}{\partial\nu}\,w
-i\left[ V \left(-\frac{1}{\partial/\partial X}
\frac{\partial}{\partial\mu}-i\,\frac{\nu}{2}
\frac{\partial}{\partial X}\right) \right.&& \nonumber\\
- \left. V\left(-\frac{1}{\partial/\partial X}
\frac{\partial}{\partial\mu}+i\,\frac{\nu}{2}
\frac{\partial}{\partial X}\right)\right]w=0\,.&&
\end{eqnarray} 
This equation is  alternative to the Schr\"odinger equation
\begin{equation}\label{news}
i\dot \Psi=H\Psi
\end{equation}
and  it can be obtained from the equation for density matrix
\begin{equation}\label{newss}
\dot \rho+i\left[H,\,\rho \right]=0,
\end{equation}
in view of the following formulas:
\begin{eqnarray}\label{eso29p}
q\,W\,(q,\,p)&\longrightarrow &-\left (\frac {\partial }{\partial X}
\right )^{-1}\,\frac {\partial }{\partial \mu}\,w\,(X,\,\mu ,\,\nu )\,,
\nonumber\\                                        
\frac {\partial }{\partial q}\,W\,(q,\,p)&\longrightarrow &
\mu \,\frac {\partial }{\partial X}\,w\,(X,\,\mu ,\,\nu )\,,\nonumber\\
&&\\
p\,W\,(q,\,p)&\longrightarrow &-\left (\frac {\partial }{\partial X}
\right )^{-1}\,\frac {\partial }{\partial \nu}\,w\,(X,\,\mu ,\,\nu )\,,
\nonumber\\                                                           
\frac {\partial }{\partial p}\,W\,(q,\,p)&\longrightarrow &
\nu \,\frac {\partial }{\partial X}\,w\,(X,\,\mu ,\,\nu )\nonumber
\end{eqnarray}                                              
and
\begin{eqnarray}\label{eso29s}                     
\frac {\partial }{\partial X}\,\rho \,(X,\,X^\prime )&\longrightarrow &
\left (\frac {1}{2}\,\frac {\partial }{\partial q}+i\,p\right )
\,W\,(q,\,p)\,,\nonumber\\
\frac {\partial }{\partial X^\prime }\,\rho \,(X,\,X^\prime )
&\longrightarrow &
\left (\frac {1}{2}\,\frac {\partial }{\partial q}-i\,p\right )
\,W\,(q,\,p)\,,\nonumber\\
&&\\
X\,\rho \,(X,\,X^\prime )&\longrightarrow &
\left (q+\frac {i}{2}\,\frac {\partial }{\partial p}\right )
\,W\,(q,\,p)\,,\nonumber\\
X^\prime \,\rho \,(X,\,X^\prime )&\longrightarrow &
\left (q-\frac {i}{2}\,\frac {\partial }{\partial p}\right )
\,W\,(q,\,p)\,.\nonumber
\end{eqnarray}
Equation~\ref{FPeq})  can be considered as a Fokker--Planck-like equation 
of classical probability theory.
The measurable position is a cyclic variable for the evolution equation.

In order to compare the classical and quantum evolution equations, let us
rewrite the quantum evolution equation~\ref{FPeq}) in the form of a series,
\begin{equation}\label{csm31}
\dot w-\mu \,\frac {\partial w}{\partial \nu }+2\,\sum _{n=0}^\infty
\frac {V^{2n+1}\left (\hat q\right )}{(2n+1)!}\left (\frac {\nu }{2}\,
\frac {\partial}{\partial X}\right )^{2n+1}(-1)^{n+1}w=0\,.
\end{equation}
Here
\begin{equation}\label{csm32}
V^{2n+1}\left( \hat q\right )=\frac {\partial ^{2n+1}V}{\partial q^{2n+1}}
\left (q=\hat q\right ),
\end{equation}
where the operator $\hat q$ is given by Eq.~\ref{csm25}).   

Equation~\ref{csm32}) can also be presented in the form
\begin{equation}\label{csm33}
\dot w-\mu \frac {\partial w}{\partial \nu }-\frac {\partial V}{\partial q}
\left (\hat q\right )\nu \frac {\partial }{\partial X}\,w+2\sum _{n=1}^\infty
\frac {V^{2n+1}\left (\hat q\right)}{(2n+1)!}\left (\frac {\nu }{2}\frac
{\partial }{\partial X}\right )^{2n+1}(-1)^{n+1}w=0.
\end{equation}
The three first terms  give the $\hbar\rightarrow 0$ classical Boltzman 
equation. 
It is important that both classical and quantum evolution equations are 
written for the same function $w\left(X,\,\mu,\,\nu\right)$.

Let us rewrite, following~\cite{ManLaser}, the Schr\"odinger equation for 
the stationary state density matrix $\rho _E$ of the quantum system 
with Hamiltonian~\ref{HV})
\begin{equation}\label{new1}
H\rho _E=\rho _EH=E\rho _E
\end{equation} 
in terms of the time-independent marginal distribution 
$w_E\left(X,\,\mu,\,\nu \right)$ of the squeezed and 
rotated quadrature introduced in~\cite{Mancini1}.
We have
\begin{eqnarray}\label{new2}
\frac {1}{2}\left (\frac{\partial}{\partial X}\right)^{-2}
\frac{\partial ^2}{\partial \nu ^2}\,w_E-\frac {1}{8}\,\mu ^2
\frac{\partial ^2}{\partial X^2}\,w_E&&\nonumber\\
+\mbox {Re}\,V\left [\frac {i}{2}\,\nu \,\frac{\partial}{\partial X}     
-\left(\frac{\partial}{\partial X}\right )^{-1}     
\frac{\partial}{\partial \mu}\right ]w_E=E\,w_E\,.&&
\end{eqnarray}                                            
The positive marginal distribution (eigendistribution) satisfies this 
eigenvalue equation and also the equation
\begin{equation}\label{new3}
-\mu \,\frac{\partial }{\partial \nu }\,w_E=2\,\mbox{Im}\,V\,
\left [\frac {i}{2}\,\nu \,\frac{\partial}{\partial X}     
-\left(\frac{\partial}{\partial X}\right )^{-1}     
\,\frac{\partial}{\partial\mu}\right ]\,w_E\,.
\end{equation}                                            
Equation~\ref{new3}) follows from the evolution equation~\ref{FPeq}) 
for the marginal distribution of the quantum system (see 
Ref.~\cite{Mancini2}\,),
if the marginal distribution does not depend on time. Thus, the normalized
marginal distributions of stationary states of quantum systems satisfy the
system of two equations~\ref{new2}) and \ref{new3}).

We consider an example of the quantum harmonic oscillator since it is one
of the most important quantum systems. For this case, using the Hamiltonian
\begin{equation}\label{new4}
H=\frac {\hat p^2}{2}+\frac {\hat q^2}{2}\,,
\end{equation}
we reduce Eq.~\ref{FPeq}) to the following one 
(in view of Ref.~\cite{Mancini2}\,):
\begin{equation}\label{TIE}
\dot w-\mu \,\frac{\partial}{\partial\nu}\,w+\,\nu \,
\frac{\partial}{\partial\mu}\,w=0\,.
\end{equation}
The marginal distribution of the oscillator's ground state is
\begin{equation}\label{new6}
w_0^{(\mbox {os})}\left(X,\,\mu,\,\nu \right)=
\frac {1}{\sqrt {\pi \,(\mu ^2+\nu ^2)\,}}
\,\exp \,\left [-\frac {X^2}{\mu ^2+\nu ^2}\right ]\,.
\end{equation}
The marginal distribution must be consistent with the uncertainty 
relation, i.e.,
\begin{eqnarray}\label{MUR}
&&\left[\int w\left(X,\,1,\,0\right)X^2\,dX
-\left\{\int w\left(X,\,1,\,0\right)X\,dX\right\}^2\right]\nonumber\\
&&\,\times\left[\int w\left(X,\,0,\,1\right)X^2\,dX
-\left\{\int w\left(X,\,0,\,1\right)X\,dX\right\}^2\right]\geq \frac {1}{4}\,.
\end{eqnarray}

\section*{Propagator}

\noindent

In Ref.~\cite{Mancini3}, the classical transition-probability density from an 
initial position $X^\prime $ measured at time $t=0$ in the reference
frame in the classical phase space labeled by the parameters 
$\mu ^\prime ;\,\nu ^\prime $
to the position $X$ measured at time $t$ in the reference
frame in the classical phase space labeled by the parameters $\mu;\,\nu $
was introduced. This classical transition-probability density is the 
propagator for the evolution equation~\ref{FPeq}) for the marginal 
distribution and the propagator is the kernel of the integral relation
\begin{equation}\label{til1}
w\,(X,\,\mu ,\,\nu ,\,t)=\int \Pi \,(X,\,\mu ,\,\nu ,\,X^\prime ,\,
\mu ^\prime,\,\nu ^\prime ,\,t)\,w\,(X^\prime ,\,\mu ^\prime,\,
\nu ^\prime ,\,0)\,dX^\prime \,d\mu ^\prime \,d\nu ^\prime \,.
\end{equation}
The classical propagator has a specific feature, it takes into account that
the transition probability is considered in different references frames 
in the phase space. In view of this fact, parameters of reference frames
$\mu$ and $\nu$ are present in the evolution equation. Due to this, the
equation for the propagator slightly differs  from the 
Smoluchowski--Chapman--Kolmogorov equation elaborated in the classical 
probability theory.

The classical propagator can be related to a quantum 
propagator (the Green function) for the density matrix 
$\rho \,(X,\,X^\prime ,\,t)$ in the coordinate representation.  
For a pure state with the wave function $~\Psi \,(X,\,t)\,,$ we have
\begin{equation}\label{til2}
\rho \,(X,\,X^\prime ,\,t)=\Psi \,(X,\,t)\,\Psi ^*(X^\prime ,\,t)\,.
\end{equation}
The Green function of the Schr\"odinger equation $G\,(X,\,X^\prime ,\,t)$
connects the wave functions at the initial time moment $t=0$ and at  
time $t$
\begin{equation}\label{til3}
\Psi \,(X,\,t)=\int G\,(X,\,X^\prime ,\,t)\,\Psi \,(X^\prime )\,dX^\prime \,.
\end{equation}
We have for the density matrix~\ref{til2}), in view of relation~\ref{til3}),
the  following expression:
\begin{equation}\label{til4}
\rho \,(X,\,X^\prime ,\,t)=\int K\,(X,\,X^\prime ,\,Y,\,Y^\prime ,t)
\,\rho \,(Y,\,Y^\prime ,\,t=0)\,dY\,dY^\prime \,,
\end{equation}
where the propagator $K\,(X,\,X^\prime ,\,Y,\,Y^\prime ,t)$ is expressed in
terms of the Green function (for unitary evolution)
\begin{equation}\label{til5}
K\,(X,\,X^\prime ,\,Y,\,Y^\prime ,t)=G\,(X,\,Y,\,t)\,
G^*(X^\prime ,\,Y^\prime \,,t)\,.
\end{equation}
Since the relation of the density matrix to the marginal distribution 
is known for any time $t$ (given by~\ref{til2}) and \ref{til3})\,), 
it is possible to obtain 
\begin{eqnarray}\label{til*}
&&K\,(X,\,X^\prime ,\,Z,\,Z^\prime ,\,t)=\frac {1}{(2\,\pi )^2}\,\int 
\frac {1}{|\nu ^\prime |}\,
\exp \,\left \{i\,\left (Y-\mu \,\frac {X+X^\prime}{2}\right ) \right. 
\nonumber\\
&& \left. -\,i\,\frac {Z-Z^\prime}{\nu ^\prime }\,Y^\prime 
+i\,\frac {Z^2-Z^{\prime 2}}{2\,\nu ^\prime }\,\mu ^\prime \right \}
\nonumber\\
&&\,\times \,
\Pi \,(Y,\,\mu ,\,X-X^\prime ,\,Y^\prime ,\,
\mu ^\prime,\,\nu ^\prime ,\,t)\,
d\mu \,d\mu ^\prime \,dY\,dY^\prime \,d\nu ^\prime \,.
\end{eqnarray}
Thus, given the classical propagator for the classical marginal 
distribution, the propagator for the density matrix 
is also given. Formula~\ref{til*})  can be converted.

Deriving  formula~\ref{til*}) we used the relations
\begin{eqnarray*}
W\,(q,\,p)&=&\frac {1}{2\,\pi}\,\int w\,(X,\,\mu ,\,\nu )\,\exp \,[-i\,
(\mu q +\nu p-X)]\,d\mu \,d\nu \,dX\,,\nonumber\\
\rho \,(X,\,X^\prime )&=&\frac {1}{2\,\pi}\,
\int w\,(Y,\,\mu ,\,X-X^\prime )\,\exp \,\left [i\,
\left (Y-\mu \,\frac {X+X^\prime }{2}\right )\right ]\,d\mu \,dY\,,
\nonumber
\end{eqnarray*}
and 
\begin{eqnarray}\label{tilc}
w\,(X,\,\mu ,\,\nu )&=&\frac {1}{2\,\pi |\nu|}\,\int \,\rho \,(Z,\,Z^\prime )
\nonumber\\
&&\,\times\,\exp \,\left [-i\,\frac {Z-Z^\prime }{\nu }\,
\left (X-\mu \,\frac {Z+Z^\prime }{2}\right )\right ]\,dZ\,dZ^\prime .
\end{eqnarray}
The last formulas give some relationships between the marginal distribution
$w\,(X,\,\mu ,\,\nu )$, the Wigner function, and the density matrix 
in the coordinate representation.

In Ref.~\cite{Wig32}, the Wigner function was introduced in terms of the
density matrix
\begin{equation}\label{Wig32.1}
W\left (q,\,p\right)=\int \rho \left (q+\frac {u}{2},\,q-\frac {u}{2}
\right )e^{-ipu}\,du\,,
\end{equation}
which can be rewritten as
$$
W\left (q,\,p\right)=\int \rho \left (Z,\,Z'\right )\,
\delta \left (Z-q-\frac {u}{2}\right )\,\delta \left (
Z'-q+\frac {u}{2}\right )e^{-ipu}\,du\,dZ\,dZ',
$$
or
\begin{equation}\label{Wig32.2}
W\left (q,\,p\right)=2\,\int \rho \left (Z,\,Z'\right )\,e^{-2ip(Z-a)}
\,\delta \left (Z'+Z-2q\right )\,dZ\,dZ'.
\end{equation}
Comparing formulas~\ref{Wig32.2}) and \ref{tilc}) one can conclude
that the Wigner quasidistribution function $W\left (q,\,p\right )$
and the classical probability distribution 
$w\left (X,\,\mu ,\,\nu\right ),$ the latter being a positive and 
normalized function, are obtained using similar integral transforms of
the density matrix.

The difference between the two functions is determined by the difference
in the kernels of the integral transforms. In the case of the Wigner
transform, the kernel reads
\begin{equation}\label{Wig32.3}
K_W\left (Z,\,Z',\,q,\,p\right )=2\,e^{-2ip(Z-q)}
\,\delta \left (Z'+Z-2q\right )\,.
\end{equation}
In the case of the symplectic tomography transform suggested 
in Ref.~\cite{Mancini1}, the kernel reads
\begin{equation}\label{Wig32.4}
K_M\left (Z,\,Z',\,X,\,\mu ,\,\nu \right )=\frac {1}{2\,\pi |\nu|}\,
\exp \left [-i\,\frac {Z-Z'}{\nu }\left (X-\mu -\frac {Z+Z'}{2}
\right )\right ]\,.
\end{equation}
Due to the difference of the kernels, the Wigner function takes 
negative values  and the marginal probability distribution is 
nonnegative function.
 
If one writes the classical propagator as a function of the initial time 
moment $t_1$ and the final time moment $t_2$ (i.e., $t_1\neq 0\,),$ 
relation~\ref{til1}) can be rewritten as
\begin{equation}\label{PI3}
w\,(X,\mu ,\nu ,t_2)=\int \Pi \,(X,\mu ,\nu ,X^\prime ,
\mu ^\prime,\nu ^\prime ,t_2,t_1)\,w\,(X^\prime ,\mu ^\prime,
\nu ^\prime ,t_1)\,dX^\prime \,d\mu ^\prime \,d\nu ^\prime .
\end{equation}
From the physical meaning of the classical propagator, the nonlinear
integral relation follows                                       
\begin{eqnarray}
&&\Pi \,(X,\,\mu ,\,\nu ,\,X^\prime ,\,\mu ^\prime,\,
\nu ^\prime ,\,t_2,\,t_1)=\int \Pi \,(X,\,\mu ,\,\nu ,\,X^{\prime \prime },\,
\mu ^{\prime \prime },\,\nu ^{\prime \prime },\,t_2,\,t^\prime )\nonumber\\
&&\,\times \,\Pi\,(X^{\prime \prime },\,
\mu ^{\prime \prime },\,\nu ^{\prime \prime },\,X^\prime ,\,\mu ^\prime,\,
\nu ^\prime ,\,t^\prime ,\,t_1)\,dX^{\prime \prime }\,d\mu ^{\prime \prime }
\,d\nu ^{\prime \prime }\,.\label{P14}
\end{eqnarray}                                          
This relation means that if the system is initially located at the point
$X^\prime $ at time $t_1$ in the reference frame in the phase space 
labeled by the parameters $\mu ^\prime ;\,\nu ^\prime ,$ the probability 
for the system to arrive at the point $X$ in the reference frame in 
the phase space labeled by the parameters $\mu ;\,\nu $ at time $t_2$ is
equal to the probabilities to arrive at an intermediate point 
$X^{\prime \prime }$ in the reference frame in the phase space labeled by 
the parameters $\mu ^{\prime \prime };\,\nu ^{\prime \prime }$ at time 
$t^\prime $ integrated over all the intermediate positions and all the
intermediate reference frames. 

The above integral equation~\ref{P14}) is an analog of the 
Smoluchowski--Chapman--Kolmogorov relation which in the approach 
introduced in Ref.~\cite{Mancini3} is generalized to the case of families
of conditional probabilities if different reference frames in the phase 
space (parameters $\mu $ and $\nu \,)$ are taken into account. Also the 
propagator satisfies the differential equation (see Ref.~\cite{Mancini3}\,)                           
\begin{eqnarray}
&&\frac {\partial \Pi }{\partial t_2}
-\mu \,\frac {\partial }{\partial \nu }\,\Pi
-i\left[V\left (-\frac {1}{\partial /\partial X}
\frac {\partial }{\partial \mu }-i\,\frac {\nu }{2}
\frac {\partial }{\partial X}\right ) \right. \nonumber\\
&&\left. -V\left (-\frac {1}{\partial /\partial X}
\frac {\partial }{\partial \mu }+i\,\frac {\nu }{2}
\frac {\partial }{\partial X}\right )\right ]\Pi \nonumber\\
&&=\delta \,(t_2-t_1)\,
\delta \,(X-X^\prime )\,\delta \,(\mu -\mu ^\prime )\,\delta \,(\nu -
\nu ^\prime )\,. \label{P15}
\end{eqnarray}                     
The classical propagator satisfies the initial condition
\begin{equation}\label{PI6}
\Pi \,(X,\,\mu ,\,\nu ,\,X^\prime ,\,\mu ^\prime,\,\nu ^\prime ,\,t,\,t)
=\delta \,(X-X^\prime )\,\delta \,(\mu -\mu ^\prime )\,\delta \,(\nu -
\nu ^\prime )\,.
\end{equation}                                          

The relation that could be used to express the classical propagator in 
terms of the functional integral can be also written
\begin{eqnarray}
&&\Pi \,(X,\,\mu ,\,\nu ,\,X^\prime ,\,\mu ^\prime,\,\nu ^\prime ,\,
t_{\mbox {f}},\,t_{\mbox {in}})\nonumber\\
&=&\int \prod _{k=1}^{N-1}\,\left \{\,\Pi \,
(X_{k+1},\,\mu _{k+1},\,\nu _{k+1},\,X_k,\,\mu _k,\,\nu _k,\,t_{k+1},\,t_k)
\,dX_k\,d\mu _k\,d\nu _k\right \}\,, 
\label{PI7}
\end{eqnarray}
where the time interval $t_{\mbox {f}}-t_{\mbox {in}}=N\,\tau \,;~
t_k=t_{\mbox {in}}+k\,\tau \,;~k=1,\,2,\,\ldots ,\,N.$ Taking in
relation~\ref{PI7})
the limit $\tau \rightarrow 0\,;~N\rightarrow \infty ,$ one obtains the
expression for the classical propagator in terms of the functional integral.

\section*{Quantum Transition Probabilities}

\noindent

We express quantum-transition probabilities in terms 
of an overlap integral of classical-like marginal distributions
describing the initial and final quantum states.

If the initial pure state of a quantum system is described by the 
marginal distribution
\begin{equation}\label{class1}
w_{\,\mbox {in}}=w_1\left(X,\,\mu,\,\nu \right)
\end{equation}
and the final state of the quantum system is described by the 
marginal distribution
\begin{equation}\label{class2}
w_{\,\mbox {f}}=w_2\left(X,\,\mu,\,\nu \right)\,,
\end{equation}
the probability of the quantum transition 
$P_{12}\left(1\Longrightarrow 2\right)$ can be obtained using the 
known expression for the probability in terms of an overlap integral 
of the Wigner functions~$W_1\left(q,\,p\right)$ and 
$W_2\left(q,\,p\right)$ of the initial and final states 
(see, for example, Ref.~\cite{183}\,)
\begin{equation}\label{class3}
P_{12}=\frac {1}{2\,\pi}\,\int W_1(q,\,p)\,W_2(q,\,p)\,dq\,dp\,.
\end{equation}
For the transition probability, one can obtain, in view of
relation~\ref{class3}),
the expression in terms of the marginal distributions
\begin{equation}\label{class4}
P_{12}=\int w_1(X,\,\mu,\,\nu )\,
w_2(Y,\,-\mu,\,-\nu )\,\exp \,[i\,(X+Y)]\,
\frac {d\mu \,d\nu \,dX\,dY}{2\,\pi}\,.
\end{equation}
As follows from relation~\ref{class4}), any pure normalized quantum state
is described by the marginal distribution 
$w_{\,\mbox {p}}\left(X,\,\mu,\,\nu \right),$ which satisfies 
the additional condition
\begin{equation}\label{class5}
\int \,w_{\,\mbox {p}}(X,\,\mu,\,\nu )\,
w_{\,\mbox {p}}(Y,\,-\mu,\,-\nu )\,\exp \,[i\,(X+Y)]
\,\frac {d\mu \,d\nu \,dX\,dY}{2\,\pi}=1\,.
\end{equation} 
The complex wave functions, which belong to different energy levels of
a quantum system, are orthogonal. This orthogonality condition is expressed
in terms of the classical marginal distribution as the relation
\begin{equation}\label{class6}
\int \,w_n(X,\,\mu,\,\nu )\,
w_m(Y,\,-\mu,\,-\nu )\,\exp \,[i\,(X+Y)]\,
\frac {d\mu \,d\nu \,dX\,dY}{2\,\pi}=\delta _{mn}\,,
\end{equation} 
where the labels $m,\,n$ correspond to the energy levels 
$E_m,\,E_n.$ The pure states $\mid n\rangle $ satisfy the completeness 
relation
\begin{equation}\label{class7}
\sum _n\,\mid n\rangle \langle n\mid =\widehat 1\,.
\end{equation}
This relation can be rewritten as the condition for the marginal
distributions of the pure states with the energies $E_n$
\begin{equation}\label{class8}
\sum _n\,w_n\left(X,\,\mu,\,\nu\right)=w^{(\mbox {wn})}\left(X,\,\mu,\,\nu
\right),
\end{equation}
where the distribution $w^{(\mbox {wn})}$ describes the white noise,
\begin{equation}\label{class8a}
w^{(\mbox{wn})}\left(X,\,\mu,\,\nu\right)=\int
\frac {dx\,dy\,dk}{2\,\pi}\exp\left[ik\left(\mu x+\nu y\right)-ikx-ik^2
\frac {\mu \nu}{2}\right].
\end{equation}

Thus, the classical marginal distributions describing the energy levels
of quantum systems are positive solutions to the system of 
equations~\ref{new2}) and \ref{new3}) and these solutions satisfy
the orthogonality condition~\ref{class6}) and the analog of the 
completeness relation~\ref{class8}). The distributions form an interesting 
mathematical set that differs substantially from the usual Hilbert space 
of states described by the normalized complex wave functions. 
Of course, the structure of the Hilbert space can be traced using the 
map, which connects the states expressed in terms of the density matrix
and the states expressed in terms of the marginal distribution functions.

\section*{Propagator for Systems with Quadratic Hamiltonians}

\noindent

As an example, we consider the system with the quadratic Hermitian 
Hamiltonian
\begin{equation}\label{qs1}
H=\frac {1}{2}\,(Q\,B\,Q)+C\,Q\,,
\end{equation}
where one has the vector-operator $Q=\left(p,\,q\right).$ The symmetric 
2$\times $2 matrix $B$ and real 2-vector $C$ depend on time.  
The system has linear integrals of motion (see Ref.~\cite{183,book}\,):
\begin{equation}\label{qs2}
I\,(t)=\Lambda \,(t)\,Q+\Delta \,(t)\,.
\end{equation}
Here the real symplectic 2$\times $2 matrix $\Lambda \left(t\right)$ 
and the real vector $\Delta \left(t\right)$ satisfy the equations
\begin{equation}\label{qs3}
\dot \Lambda =i\,\Lambda \,B\,\sigma _y\,,\qquad \dot \Delta =i\,\Lambda 
\,\sigma _y\,C\,,
\end{equation}
and the initial conditions
\begin{equation}\label{qs4}     
\Lambda \,(0)=1\,;\qquad \Delta \,(0)=0\,.
\end{equation}
As follows from relation~\ref{w}) and from the property of the Wigner 
function (see Ref.~\cite{183}\,), the classical propagator is
\begin{equation}\label{qs5}
\Pi \,(X,\,\mu ,\,\nu ,\,X^\prime ,\,\mu ^\prime,\,\nu ^\prime ,\,t)
=\delta \,(X-X^\prime +{\cal N}\,\Lambda ^{-1}\Delta )\,\delta 
\,({\cal N}^\prime -{\cal N}\,\Lambda ^{-1})\,,
\end{equation}                   
where the vectors ${\cal N}$ and ${\cal N}^\prime $ are
\begin{eqnarray*}
{\cal N}&=&(\nu ,\,\mu )\,,\nonumber\\
{\cal N}^\prime &=&(\nu ^\prime ,\,\mu ^\prime )\,.\nonumber
\end{eqnarray*}
For the quadratic systems without linear terms $\left(C=0\right),$ 
the classical propagator is                                                     
\begin{equation}\label{qs6}
\Pi \,(X,\,\mu ,\,\nu ,\,X^\prime ,\,\mu ^\prime,\,\nu ^\prime ,\,t)
=\delta \,(X-X^\prime )\,\delta 
\,({\cal N}^\prime -{\cal N}\,\Lambda ^{-1})\,.
\end{equation}                                                    
Thus, if one knows the linear integrals of motion, i.e., the matrix
$\Lambda \left(t\right)$ and the vector $\Delta \left(t\right)$, 
one knows the classical propagator.

For free motion with the Hamiltonian
\begin{equation}\label{qs7}
H=\frac {p^2}{2}\,,
\end{equation}
there are two linear invariants found in Ref.~\cite{183,book}         
\begin{equation}\label{qs8}
p_0(t)=p\,,\qquad q_0(t)=q-pt\,.
\end{equation}                                           
This means that $\Delta \left(t\right)=0,$ and the symplectic 
2$\times $2 matrix reads
\begin{equation}\label{qs9}
\Lambda \,(t)=\left (\begin{array}{clcr}
1&0\\
-t&1\end{array}\right )\,.
\end{equation}
Thus, we have
\begin{equation}\label{qs10}
{\cal N}\,\Lambda ^{-1}(t)=(\nu +\mu t,\,\mu )\,.
\end{equation}
Consequently, the classical propagator of free motion has the form  
\begin{equation}\label{qs11}
\Pi ^{(\mbox {f})}(X,\,\mu ,\,\nu ,\,X^\prime ,\,\mu ^\prime,\,
\nu ^\prime ,\,t)=\delta \,(X-X^\prime )\,
\delta \,(\nu ^\prime -\nu -\mu t)\,
\delta \,(\mu -\mu ^\prime )\,.
\end{equation}                                                      

For the harmonic oscillator with the Hamiltonian~\ref{new4}), linear 
invariants are known (see Ref.~\cite{183,book} \,), and the matrix 
$\Lambda \left(t\right)$ is
\begin{equation}\label{qs12}
\Lambda \,(t)=\left (\begin{array}{clcr}
\cos \,t&\sin \,t\\
-\sin \,t&\cos \,t\end{array}\right )\,.
\end{equation}
This means that for the harmonic oscillator
\begin{equation}\label{qs13}
{\cal N}\,\Lambda ^{-1}(t)=(\nu \,\cos \,t-\mu \,\sin \,t,~\nu \,\sin \,t+
\mu \,\cos \,t)\,.
\end{equation}
Consequently, the classical propagator of the harmonic oscillator is
\begin{eqnarray}\label{qs14}
&&\Pi ^{(\mbox {os})}(X,\,\mu ,\,\nu ,\,X^\prime ,\,\mu ^\prime,\,\nu ^\prime )
\nonumber\\
&=&\delta \,(X-X^\prime )\,\delta \,(\nu ^\prime -\nu \,\cos \,t+\mu \,\sin \,t)
\,\delta \,(\mu ^\prime -\nu \,\sin \,t-\mu \,\cos \,t)\,.
\end{eqnarray}                                                      

\section*{Energy Levels of the Harmonic Oscillator}

\noindent

The marginal distribution of the coherent state  of the harmonic oscillator
has the form obtained in Ref.~\cite{Mancini3}\,:
\begin{eqnarray}\label{eso10}
w_\alpha (X,\,\mu ,\,\nu )&&=[\pi\,(\mu^2+\nu ^2)]^{-1/2}\,\exp \left [
-|\alpha |^2-\frac {X^2}{\mu ^2+\nu ^2}+\frac {\alpha ^2(\nu +i\mu )^2}
{2\,(\mu ^2+\nu ^2)}\right.\nonumber\\
&&+\left. \frac {\alpha ^{*2}(\nu -i\mu )^2}{2\,(\mu ^2+\nu ^2)}
-\frac {i\sqrt 2\alpha X(\nu +i\mu )}{\mu ^2+\nu ^2}+\frac {i\sqrt 2\alpha ^* 
X(\nu -i\mu )}{\mu ^2+\nu ^2}\right ] .                                                                  
\end{eqnarray}
The eigendistribution function for the energy level of the harmonic oscillator 
satisfies the eigenvalue equation                          
\begin{eqnarray}\label{eso14}
&&\left \{\frac {1}{2}\left [\left (\frac {\partial}{\partial \nu }
\right )^2+\left (\frac {\partial}{\partial \mu }\right )^2\right ]
\left (\frac {\partial}{\partial X}\right )^{-2}-\frac {1}{8}\,
(\mu ^2+\nu ^2)\left (
\frac {\partial}{\partial X}\right )^2\right \}\,w_E(X,\,\mu ,\,\nu )   
\nonumber\\
&&=E\,w_E(X,\,\mu ,\,\nu )\,.
\end{eqnarray}                                                   
This equation can be rewritten for the Fourier component of the marginal
distribution                                          
\begin{equation}\label{eso15}
\widetilde w_E(k,\,\mu ,\,\nu )=\frac {1}{2\,\pi }\,\int \,
w_E(X,\,\mu ,\,\nu )\,
\exp \,(-ikX)\,dX
\end{equation}                                        
in the form                                             
\begin{equation}\label{eso16}
\left \{-\frac {1}{2\,k^2}\left [\left (\frac {\partial}
{\partial \nu }\right )^2+\left (\frac {\partial}{\partial \mu }\right )^2
\right ]+\frac {1}{8}\,k^2(\mu ^2+\nu ^2)\right \}\,\widetilde
w_E(k,\mu ,\nu )   
=E\,\widetilde w_E(k,\mu ,\nu ).
\end{equation}                                          
Since the marginal distribution of the stationary state of the harmonic
oscillator must satisfy the stationarity condition found in Ref.~\cite{ManConf},
\begin{equation}\label{eso17}
\left (\mu \,\frac {\partial}{\partial \nu }-\nu \,\frac {\partial}
{\partial \mu }\right )\,w_E(X,\,\mu ,\,\nu )=0\,,   
\end{equation}                                      
Eq.~\ref{eso16}) is equivalent to the equation for axially symmetric wave
functions of a two-mode harmonic oscillator with mass $m=k^2,$ 
frequency $\omega =1/2,$ and angular momentum $M=0.$ 
The wave function corresponding to zero angular
momentum is expressed in terms of the Laguerre polynomials 
\begin{equation}\label{eso18}
\widetilde w_n(k,\,\mu ,\,\nu )=\frac {1}{2\,\pi }\,\exp \,
\left [-\frac {k^2(\mu ^2
+\nu ^2)}{4}\right ]\,L_n\left (\frac {k^2\mu ^2+k^2\nu ^2}{2}\right )\,.
\end{equation}                                            
The main quantum number $n$ of the one-dimensional harmonic oscillator under
discussion is equal to the integer radial quantum number $n_r$ of the
artificial two-mode oscillator                           
\begin{equation}\label{eso19}
n=n_r\,,\qquad n_r=0,\,1,\,2,\,\ldots \,.  
\end{equation}                                           
The energy level of the artificial symmetric two-mode oscillator labeled 
by the radial quantum number $n_r$ and the angular momentum $M$ as
\begin{equation}\label{eso20}
E_{n_r,\,M}=\omega \,(|M|+1+2\,n_r)
\end{equation}                 
for $~\omega =1/2\,,~M=0\,,~n_r=n~$ gives exactly the spectrum of the
one-dimensional oscillator
\begin{equation}\label{eso21}
E_n=E_{n_r,\,M}=n+\frac {1}{2}\,,\qquad n=0,\,1,\,2,\,\ldots   
\end{equation}                                      
To find the marginal distribution, we have to calculate
\begin{equation}\label{eso22}
w_n(X,\,\mu ,\,\nu )=\frac {1}{2\,\pi }\,\int \,\exp \,\left [-
\frac {k^2(\mu ^2+\nu ^2)}{4}+ikX\right ]\,L_n\left (
\frac {k^2\mu ^2+k^2\nu ^2}{2}\right )\,dk\,.
\end{equation}                              
In view of the integral
\begin{eqnarray}\label{eso23}
&&\frac {1}{2\,\pi }\,\int _{-\infty }^\infty \,\exp \,\left (-
\frac {k^2}{4}+ikX\right )\,L_n\left (
\frac {k^2}{2}\right )\,dk\nonumber\\
&&=\pi ^{-1/2}\,2^{-n}\,(n!)^{-1}\,\exp \,(-X^2)
\,H_n^2(X)\,,
\end{eqnarray}         
one obtains the marginal distribution 
\begin{eqnarray}\label{eso13}
w_n(X,\,\mu ,\,\nu )&=&[\pi\,(\mu^2+\nu ^2)]^{-1/2}
\,2^{-n}\,(n!)^{-1}\nonumber\\
&&\,\times\,\exp \,\left (-\frac {X^2}{\mu ^2+\nu ^2}\right )\,
H_n^2\left (\frac {X}{\sqrt {\mu ^2+\nu ^2}}\right )\,,
\end{eqnarray}
It is worth noting that the normalization condition for the marginal
distribution $w_n(X,\,\mu ,\,\nu )$ implies the condition for the Fourier
component                              
\begin{equation}\label{eso24}
\int \,\widetilde w_n(k,\,\mu ,\,\nu )\,\exp \,(ikX)\,dk\,dX=2\,\pi 
\widetilde w_n(k=0,\,\mu ,\,\nu )=1\,. 
\end{equation}
We take solutions~\ref{eso18}) without using the normalization condition 
in terms of the variables $\mu $ and $\nu $ of the artificial two-mode 
oscillator, but using the normalization condition of the marginal 
distribution in terms of the variable $X$ and the corresponding property 
of its Fourier component.

\section*{Tomography of  Spin States}
 
\noindent

Let us introduce the  probability
distribution for the spin projection in a given direction considered
in a rotated reference frame. 
For arbitrary values of spin, let the spin state have the density matrix 
\begin{equation}\label{d1}
\rho _{mm'}^{(j)}=\langle jm\mid \hat \rho^{(j)}\mid jm'\rangle \,,
\qquad m=-j,-j+1,\ldots,j-1,j\,, 
\end{equation}
where
\begin{eqnarray}\label{eq.1}
\hat j_3 \mid j m \rangle& =& m \mid j m \rangle\,,\nonumber \\
&&\\
\hat j^2 \mid j m\rangle& =& j(j+1) \mid j m\rangle\,,\nonumber
\end{eqnarray}
and
\begin{equation}\label{eq.2}
\hat\rho^{(j)}=\sum_{m=-j}^j \sum_{m'=-j}^j \rho_{m
m'}^{(j)}\mid j m\rangle\langle j m'\mid \,.
\end{equation}
The operator $\rho^{(j)}$ is the density operator of the state under 
discussion. The diagonal elements of the density matrix determine the 
positive probability distribution
\begin{equation}\label{eq.3}
\rho_{m m }^{(j)} = w_0(m)\,,
\end{equation}
which is normalized,
\begin{equation}\label{eq.4}
\sum_{m=-j}^j w_0(m)=1\,.
\end{equation}
In Refs.~\cite{Mancini4,Tom5,Tom6}, a general group construction of tomographic
schemes was discussed, and this scheme was also used for spin tomography in 
Refs.~\cite{Dodspin,JETF}. The idea is to consider the diagonal elements of
the density matrix in another reference frame. The density matrix in another 
reference frame reads
\begin{equation}\label{d2}
\rho^{(j)}_{m_1\,m_2}=\left ({\cal D}\rho {\cal D}^\dagger 
\right )_{m_1\,m_2}\,.
\end{equation}
Here the unitary rotation transform ${\cal D}$ depends on the Euler angles
$\alpha ,\,\beta ,\,\gamma \,$ and, by definition, the diagonal matrix 
elements of the density matrix yield the positive normalized probability 
distribution. For the diagonal elements of the density matrix~\ref{d2}),
$$m_1=m_2\,.$$
We introduce new notation and rewrite equality~\ref{d2}) for
$m_1=m_2$ in the form
\begin{equation}\label{eq.5}
\widetilde w\left(m_1,\alpha,\beta,\gamma\right) = 
\sum^j_{m_1'=-j}\,\sum^j_{m_2'=-j}
\,D_{m_1m_1'}^{(j)}(\alpha, \beta,\gamma)\,
\rho^{(j)}_{m_1'm_2'}\,D_{m_1m_2'}^{(j)\ast}(\alpha, \beta, \gamma)\,.
\end{equation}
Here the matrix elements 
$D_{m_1\,m_1'}^{(j)}\left (\alpha, \beta, \gamma \right)$~(the Wigner 
function) are the matrix elements of the rotation-group representation
\begin{equation}\label{eq.6}
D^{(j)}_{m'm}(\alpha,\beta,\gamma)=e^{i m'\gamma}\,d_{m'm}^{(j)}(\beta)\,
e^{i m \alpha}\,,
\end{equation}
where 
\begin{eqnarray}\label{eq.7}
d_{m'm}^{(j)}(\beta)&= &\left[\frac{(j+m')!(j-m')!}{(j+m)!(j-m)!}\right]^{1/2}
\left(\cos\,\frac{\beta}{2}\right)^{m'+m} \left(\sin\,
\frac{\beta}{2}\right)^{m'-m}\nonumber\\
&&\,\times \, P_{j-m'}^{(m'-m,m'+m)}(\cos\,\beta)\,,
\end{eqnarray}
and $P_n^{(a,b)}(x)$ is the Jacobi polynomial.

Since
\begin{equation}\label{eq.8}
D_{m'm}^{(j)\ast}(\alpha,\beta,\gamma)=(-1)^{m'-m}D_{-m'
-m}^{(j)}(\alpha,\beta,\gamma)\,,
\end{equation}
the marginal distribution depends only on two angles, $\alpha$ and 
$\beta .$

Thus, let us denote
\begin{equation}\label{eq.9}
w\left(m_1,\alpha,\beta\right)=\widetilde w\left(m_1,\alpha,\beta,\gamma
\right)\,,
\end{equation}
which satisfies
\begin{equation}\label{eq.10}
\sum_{m_1=-j}^jw\left(m_1,\alpha,\beta\right)=1\,.
\end{equation}
For a spin-1/2 state with spin projection $+1/2$ and wave function
$$
\psi_{+1/2}=\pmatrix{1 \cr 0}\,,
$$
or with density matrix
$$
\rho_+=\pmatrix{1 & 0 \cr 0 & 0}\,,
$$
the marginal distribution is equal to
\begin{equation}\label{d3}
w\left (\frac {1}{2},\alpha,\beta\right )=\cos ^2\frac {\beta}{2}
\qquad \mbox {for} \qquad m_1=+\frac {1}{2}\,,
\end{equation}
and, correspondingly, 
\begin{equation}\label{d4}
w\left (-\frac {1}{2},\alpha,\beta\right )=\sin ^2\frac {\beta}{2}
\qquad \mbox {for} \qquad m_1=-\frac {1}{2}\,.
\end{equation}
In Ref.~\cite{JETF}, by using the properties of the Wigner function and the 
Clebsch--Gordan coefficients, formula~\ref{eq.5}) was inverted and the
density matrix was expressed in terms of the marginal distribution
\begin{eqnarray}\label{eq.17}
&&(-1)^{m_2'}\sum_{j_3=0}^{2j}\,\sum_{m_3=-j_3}^{j_3}\,(2j_3+1)^2\sum_{m_1
=-j}^{j} \int (-1)^{m_1} w\left(m_1,\alpha,\beta\right)\, D_{0 m_3}^{(j_3)}
(\alpha,\beta,\gamma)\nonumber\\
&&\,\times\, \pmatrix{j&j&j_3\cr m_1&-m_1&0}\pmatrix{j&j&j_3\cr m_1'&-m_2'&m_3}
\frac{d\omega}{8\pi^2}=\rho_{m_1'm_2'}^{(j)}.  \end{eqnarray}
Here $m,m^{'}=-j,-j+1,\ldots ,j$
 and one integrates over rotation parameters
$\alpha ,\,\beta ,\,\gamma $.
 
To derive formula~\ref{eq.17}), we used the known property of the Wigner 
function: 
\begin{eqnarray}\label{Zheq.11}
\int D_{m_1'm_1}^{(j_1)}(\omega)D_{m_2'm_2}^{(j_2)}(\omega)
D_{m_3'm_3}^{(j_3)}(\omega)
\frac{d\omega}{8\pi^2} &=&\pmatrix{j_1&j_2&j_3\cr m_1'&m_2'&m_3'}
\nonumber\\
&&\,\times \pmatrix{j_1&j_2&j_3\cr m_1&m_2&m_3},
\end{eqnarray}
where
\begin{equation}\label{Zheq.12}
 \int d\omega =\int _0^{2\pi}d\alpha \int _0^{\pi} \sin \,\beta \,d\beta
\int _0^{2\pi}d\gamma \,,
\end{equation}
along with the orthogonality property of 3$j$-symbols:
$$
(2j+1)\sum_{m_1=-j_1}^{j_1}\,\sum_{m_2=-j_2}^{j_2} 
\pmatrix{j_1&j_2&j\cr m_1&m_2&-m}
\pmatrix{j_1&j_2&j'\cr m_1&m_2&-m'}=\delta_{jj'}\,\delta_{mm'},
$$
$$
\sum_{j=\mid j_1-j_2\mid}^{j_1+j_2}\,\sum_{m=-j}^{j}(2j+1)
\pmatrix{j_1&j_2&j\cr m_1 &m_2& -m}\pmatrix{j_1&j_2&j\cr m_1'&m_2'&
-m}=\delta_{m_1 m_1'}\,\delta_{m_2 m_2'}.
$$
Formula~\ref{eq.17}), being the inverse 
of~\ref{eq.5}),  is an analog of the Radon transform for spin states.
{\it Given a measurable marginal distribution for arbitrary spin,
one can reconstruct the state density matrix by means of this relation.}

The results obtained enable one to measure the spin state by 
measuring a spin projection on a given axis. One obtains the experimental 
probability-distribution function which depends on two angles determining 
the axis. Using the relationship between the probability distribution and the
state density matrix, one reconstructs all the information about the 
quantum spin state. This means that the probability distribution can be used 
instead of complex spinors and density matrices for the spin-state 
description since it contains complete (even overcomplete) information
on the state.

\section*{Quantum Measurements and Collapse of Wave Function}
 
\noindent

We will review the discussion of quantum measurements done
in Refs.~\cite{Mancini2,Mancini3,Bregenz96}. It is known (see, for example, 
Refs.~\cite{WZ,Bell}\,) 
that quantum  mechanics is problematic in the sense that it is incomplete and 
needs the notion of a classical device measuring quantum observables as an 
important ingredient of the theory. Due to this, one accepts that there 
exist two worlds: the classical one and the quantum one. In the classical 
world, the measurements of classical observables are produced by 
classical devices. In the framework of standard theory, in the quantum 
world the measurements of quantum observables are produced by 
classical devices, too. Due to this, the theory of quantum 
measurements is considered as something very specifically different from 
classical measurements.

It is psycologically accepted that to understand the physical meaning of a 
measurement in the classical world is much easier than to understand the 
physical meaning of analogous measurement in the quantum world. 

As was pointed out in Refs.~\cite{Mancini2,Mancini3}, all the roots of the 
difficulties of quantum measurements are present in classical measurements, 
as well. Using the relations of the quantum states in the standard 
representation and in the classical one (described by classical distributions),
one can conclude that complete information on a quantum state is obtained from 
purely classical measurements of the position of a particle made 
by classical devices in each reference frame of an ensemble of classical 
reference frames, which are scaled and rotated in the classical phase space.

These measurements do not need any quantum language if we know how to 
produce, in the classical world (using the notion of classical position and 
momentum), reference frames in the classical phase space differing from each 
other by rotation and scaling of the axis of the reference frame and how to 
measure only the position of the particle from the viewpoint of these 
different reference frames. So, knowing how to obtain the classical marginal
distribution function $w\left(X,\,\mu ,\,\nu \right)$ which depends on the 
parameters $\mu ;\,\nu ,$ labeling each reference frame in the classical 
phase space, we reconstruct the quantum density operator. 

Thus, we avoid the paradox of the quantum world which requires for its 
explanation measurements by a classical apparatus accepted in the framework of  
standard treatment of quantum mechanics. But the difficulties of the quantum 
approach are present, since we need to understand better the procedure of
measurement in a rotated reference frame in the phase space of the classical  
system. The problem of wave function collapse~\cite{WZ,Bell} reduces to the 
problem of a reduction of the probability distribution which occurs as soon 
as we ``pick'' a classical value of the classical random observable in the 
classical framework of~\cite{Mancini2,Mancini3}.
This means that we ``solved'' the paradox of the  wave function collapse 
reducing it to the problem of  standard measurement of a classical random 
variable used in the probability theory. 

The approach developed in~\cite{Mancini2,Mancini3} enables one to transform 
such an unpleasant problem of standard quantum mechanics as the need of a 
classical device and the reduction of wave packets into the standard problem of
classical measurements of classical random variables. In fact, this means that 
the problem of classical measurements is as difficult as the problem of
quantum measurements. An important analogy with methodology of special
relativity arises:~ It turns out that it is necessary to introduce a
consideration of events in the set of moving reference frames in space--time 
in order to explain relativistic effects, and it is necessary to 
introduce a consideration of events in the set of rotated and scaled 
reference frames in the phase space in order to explain the nonrelativistic 
quantum mechanics in terms of only classical concepts of classical fluctuation
theory. But these reference frames are the reference frames in the phase space 
(not in space--time). Possibly, a combination of these two approaches  can be 
generalized to give a classical description of relativistic
quantum mechanics.   

One can conclude that the stationary states of quantum systems
(for example, of a harmonic oscillator) can be obtained using classical-like 
alternative equations to the Schr\"odinger equations. A new type of 
eigenvalue problems for real positive marginal distributions is formulated. 
The analogs of orthogonality and completeness relations for the wave 
functions are formulated in terms of conditions for the marginal 
distributions as well as the transition probabilities among the energy levels. 
The criterion for determining the pure states of the quantum system is given 
in terms of the classical marginal distribution.
                                                    
Thus, using the marginal distribution one can formulate the standard 
quantum mechanics without the complex wave function and density matrix. 
But the position distributions in an ensemble of classical reference frames 
in  the phase space play an important role.

It should be pointed out that in the standard formulation of quantum 
mechanics
there exist different representations such as the coordinate representation,
the momentum representation, etc. The counterpart of this variety in the 
classical
formulation is related to different tomography schemes like optical
tomography~\cite{VogRi,Raymer}, symplectic tomography~\cite{Mancini3,Mancini1},
and photon number tomography~\cite{Mancini4,WalVog,Wod}.
The photon number tomography uses the marginal distribution of a discrete
variable, which corresponds to a number representation. Just as different
representations in the standard formulation of quantum mechanics are related by
some transformations in the Hilbert space of states, the marginal distributions
of different tomography schemes can be transformed into each other. This
transformation consists of two steps. First, one makes a map of the marginal
distribution (in one of the tomography schemes) onto the Wigner function and 
then one makes another map (of the different tomography scheme) of this Wigner
function onto the corresponding marginal distribution.  

The construction introduced  in spirit is similar to the Moyal 
approach~\cite{Moy} which considers quantum mechanics as a statistical theory.
But in Ref.~\cite{Moy}, the quantum state was described by a quasidistribution
function in the phase space that is identical to the Wigner function.
Thus, the ``negative probabilities'' to find the system in some
domain in the phase space is an unavoidable feature of the Moyal approach. 
The density matrix was introduced in Ref.~\cite{L.L.,Ne}.
In the framework of  the new formulation of quantum mechanics,  the 
density matrix  is not mandatory to be used.

In the introduced formulation of quantum mechanics, only positive
probabilities of  the measurable position in an ensemble of reference 
frames in the 
phase space of the system is used. It is remarkable that in the positive
probability representation the states in quantum mechanics are described 
identically with the states in classical statistical mechanics if one uses
the positive marginal distribution $w\left (X,\,\mu ,\,\nu \right )\,$
(though the sets of the distributions in the classical and quantum cases are 
different).  

The difference between classical statistical mechanics and quantum mechanics
in the formulation introduced appears in the dynamics of the marginal 
distributions, since in quantum mechanics the evolution equation for the 
positive probability distribution has a different form from that in 
the classical case. It is remarkable that the relations of the propagators
(conditional probabilities) in the phase space representation and in the 
probability 
representation are described by the same formula both in 
classical statistical mechanics and in quantum mechanics.

For linear systems (oscillators), the propagators in classical stastistical
mechanics and in quantum mechanics coincide. The difference for these systems 
in the quantum and classical cases is due to the fact that not all positive
probability distributions $w\left (X,\,\mu ,\,\nu \right )$ are realized for
classical systems. Also not all marginal distributions are admissible  in 
the quantum case, but  only  which satisfy  uncertainty relations. 

We have demonstrated that spin states and states of a trapped 
ion~\cite{Olgaphl} can be
described by measurable positive probability distributions. This implies
that quantum-mechanical systems   can  be considered in the framework of
the same formalism of probability theory as classical statistical
systems. Thus, the known results of classical probability theory
can be applied to the study of quantum states. For example, the central limit 
theorem can be used for describing multimode systems.
The approach developed can be elaborated for solving many problems of quantum 
optics~\cite{ELAF95,WunMa,SchraMa} and quantum computing.
It is important to study  the Schr\"odinger uncertainty relation~\cite{Schr30}  
in the framework of the new approach.
Linear integrals of motion for quadratic systems~\cite{MM70,Tri70} are 
useful to obtain the propagator of the new evolution equation for the 
marginal distribution~\cite{ROSAPhysRev}. 
The new approach can be also applied to study nonlinear coherent 
states~\cite{Guad,F-Scr,FilVog}.

We have shown that quantum mechanics can be formulated without wave 
function and density matrix using the tomographic probability 
representation. The general approach to the tomographic map 
and relations among different  tomography schemes are discussed in 
Refs.~\cite{MANModOpt,MANSemOpt}. The tomography of spin states  
for two particles is described in Ref.~\cite{ANDREEVJETP}.  A review of 
the new approach to quantum mechanics is given in Ref.~\cite{ANDREEVJRLR}. 
The generalization of the metod to the case of the field theory  
one can find in 
Ref.~\cite{ROSAPhysLett}. One can conclude that the problem of 
formulation of quantum theory  using only probabilities both for 
continuous and discrete  observables has the solution in the framework 
of the tomographic probability representation of quantum mechanics.
  
\section*{Acknowledgments}

\noindent

The author  thanks the Organizers of the  XXXI Latin-American School of 
Physics  for invitation to give a course of lectures and El Colegio 
Nacional for kind hospitality.

This  study has been partially supported by the Russian Foundation for 
Basic Research under Project~No.~17222.

\end{document}